\documentclass{article} 
\usepackage{iclr2025_conference,times}

\usepackage{amsmath,amsfonts,bm}

\def\eqref#1{equation~\ref{#1}}

\def\1{\bm{1}}

\DeclareMathAlphabet{\mathsfit}{\encodingdefault}{\sfdefault}{m}{sl}
\SetMathAlphabet{\mathsfit}{bold}{\encodingdefault}{\sfdefault}{bx}{n}

\usepackage{hyperref}
\usepackage{url}
\usepackage{soul}
\usepackage{xspace}
\usepackage{graphicx}
\usepackage{booktabs}
\usepackage{multicol}
\usepackage{multirow}
\usepackage{makecell}
\usepackage[table]{xcolor}
\usepackage{paralist}
\usepackage{tcolorbox}
\usepackage{float}
\usepackage{adjustbox}
\usepackage{pgfplots}

\newcommand{\ie}{\emph{i.e.,}\xspace}
\newcommand{\eg}{\emph{e.g.,}\xspace}

\title{On Pretraining \\ for Project-Level Code Completion}

\author{Maksim Sapronov, Evgeniy Glukhov
\\
JetBrains Research\\
\texttt{\{name.last\_name\}@jetbrains.com} 
}

\iclrfinalcopy

\begin{document}

\maketitle

\begin{abstract}
\textit{Repository-level pretraining} is commonly used to enable large language models for code to leverage codebase-wide context. This enhances their ability to generate accurate and context-aware code completions. In this work, we investigate how different repository-processing strategies affect in-context learning in OpenCoder, a 1.5B-parameter model. We extend its context window from 4,096 to 16,384 tokens by training on additional 1B tokens of curated repository-level data. Despite relying on a smaller dataset than competing models (which often use hundreds of billions of tokens), our model achieves comparable performance on the Long Code Arena benchmark. We find that various repository-processing techniques yield similarly strong results, with the primary gain coming from adapting to a new rotary positional embedding (RoPE) scaling parameter. Finally, we show that a simpler file-level training approach at the original sequence length remains highly effective, opening up repository-level code completion research to settings with more constrained data and compute resources.
\end{abstract}

\section{Introduction and Motivation}

Large Language Models (LLMs) trained on source code, commonly known as Code LLMs, have demonstrated impressive capabilities on a variety of code-related tasks \citep{codexglue, swe-bench, LLM-SE-review, CodeLLM-survey}. Traditionally, these models are pretrained on individual files, effectively capturing local context but often missing broader, project-level information. To address this limitation, several recent works have incorporated a \textit{repository-level pretraining} phase, \ie a stage of pretraining during which the model gets training examples from entire repositories to learn context spanning multiple files, shared dependencies, and cohesive development patterns. For example, models such as DeepSeek Coder, Starcoder 2, Qwen2.5 Coder and CodeGemma \citep{deepseek-coder, starcoder2, qwen2p5coder, codegemma} incorporate repository-level pretraining to extend their context windows and capture cross-file relationships. Beyond repository-level pretraining, other techniques have been investigated \citep{long-coder, repocoder, hirope}.

While repository-level pretraining enhances long-context capabilities, it also introduces significant challenges. Firstly, it requires huge amounts of data, \eg Qwen2.5 Coder’s repository-level pretraining leverages approximately 300B tokens of repository data. Secondly, long sequences can strain computational resources due to the quadratic complexity of traditional transformer architecture. Recent advances in efficient attention mechanisms (for example, Flash Attention and Ring Attention \citep{flash-attention-2, ring-attention}) have enabled training with context lengths typically in the tens of thousands of tokens, and even millions of tokens for smaller models. However, effective utilization of repository-level information remains challenging both for training and inference \citep{cocomic, repobench, cross-code-eval, better-context-better-completion}.

In this work, we focus on a single line repository-level code completion and study context extension pretraining for various repository preprocessing approaches. Following the Long Code Arena benchmark \citep{LCA} terminology, we evaluate the impact of different \textit{context composers}, \ie processors that transform repository files into context strings. Our approach builds on the OpenCoder base model \citep{opencoder}, originally configured with a 4,096 (4K) context window, by training on repository-level input sequences of up to 16,384 (16K) tokens. This extension results in significantly improved performance on 16K token sequences compared to the initial configuration.

We assess our methods using the Project-Level Code Completion task from the Long Code Arena benchmark, which effectively estimates a model’s ability to handle cross-file dependencies in realistic settings. By isolating the impact of repository-level pretraining and comparing different context composer strategies, our study provides practical insights for enhancing long-context code completion performance.

The main contributions of this paper are:
\begin{enumerate}
    \item We boost the project-level code completion performance of OpenCoder 1.5B to state-of-the-art levels using only 1B tokens of training data;
    \item Our experiments reveal that the choice of context composer during pretraining has only a marginal impact on final model quality, with performance scores ranging from 45.2 to 48.8 (out of 100) on the chosen metric.
\end{enumerate}

\section{Experiment Design}
We start this section with a description of the data sourcing and preparation steps, then explain our training setup, and finally we detail our evaluation strategy. In addition, we discuss the role of context composers --- repository processing functions, and their distinct modes for the training and evaluation phases.

\subsection{Training Data}
\label{paper:data}
To collect the training data, we follow the approach from the Long Code Arena benchmark, see \citet{LCA} for more details. Starting with open-source GitHub repositories in Python and then traverse the Git commit history for each repository to extract repository data. Filtering process is described in \ref{appendix:data-filtering}

The repository data for each commit consists of two elements:
\begin{inparaenum}[(1)]
    \item \textit{repository snapshot} --- a context source with contents of all code and text files before the commit;
    \item \textit{completion files} --- list of files to perform completion on with contents of all \texttt{.py} files added in that commit.
\end{inparaenum}

The resulting \textit{raw repositories dataset} contains 1,640 repositories, 160,801 commits, and 361,052 completion files. 
The total number of characters in completion files is 1.7B, and in repository snapshot files --- 4.8T.

To get a \textit{context string} from the repository data, we apply a \textit{context composer} to a repository snapshot. A \textit{context composer} is a repository processor that \begin{inparaenum}[(1)]
\item sorts a subset of files (or file chunks) from the repository snapshot by relevance (based on specified criteria),
\item retrieves the most relevant ones that fit within the context window, and 
\item concatenates them into a single string with the most relevant file at the end.
\end{inparaenum}

For each context composer in the list provided in Appendix \ref{appendix:context-composers}, we prepare the \textit{composed dataset} from the raw repositories dataset which consists of two columns:
\begin{inparaenum}[(1)]
    \item \textit{completion file} --- one file from the completion files;
    \item \textit{composed context} --- string with the result of the context composer.
\end{inparaenum}

Of the various context composers used in our experiments, we highlight the following two for clarity and conciseness.
\begin{enumerate}
    \item \textbf{File-level} — Produces an empty context.
    
    \item \textbf{Path Distance \texttt{.py}} — The context is built solely from .py files, sorted in descending order by their path distance from the completion file. For files with the same distance, a secondary sort uses the Intersection over Union (IoU) score of their matching lines.
\end{enumerate}

Note that rows from the raw repositories dataset can produce multiple rows of the composed dataset with one row for each completion file.

\subsection{Training}
For each context composer, we pretrain OpenCoder 1.5B model \citep{opencoder} on the corresponding composed dataset with a context window size of 16,384 tokens.

In our \textit{training mode} for the context composer, we aim to include as many files as possible in the context string. To achieve this, we truncate both the context string and the completion file, ensuring that the context-to-completion token ratio is at least $3:1$. For more details, see Appendix \ref{appendix:training-cococo}.

To extend the model context window size, we change RoPE's base frequency $\theta$ from 10,000 to 500,000 following the ABF approach \citep{RoPE-ABF}; our focus is on this method, although alternative approaches exist \citep{positional-interpolation, yarn, rope-extensions, rope-scaling-laws}.

To evaluate models after training on approximately 1 billion tokens, and in accordance with our training hyperparameters (see Appendix \ref{appendix:training-hyperparams}), we save the model’s weights at the 512th optimization step, referring to this saved state as a \textit{checkpoint}.

\subsection{Evaluation}
Our evaluation setup is based on the \textit{large context} dataset, which is a part of the Project-level code completion task from the Long Code Arena dataset (LCA-large) \citep{LCA}. The task is to write the next line of code based on the file prefix and the repository snapshot, with the evaluation metric being Exact Match (percentage of correct answers). Additionally, each line has one of six categories that corresponds to various scenarios of project cross-file dependencies. We use categories \textit{infile} and \textit{inproject}, \ie a completion line that contains an API declared in the completion file or in repository snapshot files. These two categories indicate in-context learning capabilities the best out of six, since they contain more project-specific information.

We evaluate each checkpoint on \textit{infile} and \textit{inproject} categories for two different context composers in the \textit{evaluation mode}: 
\begin{inparaenum}[(1)]
\item \textbf{FL-4K}: File-Level composer with maximum sequence length 4K tokens, and
\item \textbf{PD-16K}: Path Distance \texttt{.py} composer with maximum sequence length 16K tokens.
\end{inparaenum} 
Moreover, we calculate \textbf{RCB}: repository-context boost, \ie the difference between scores for the PD-16K and FL-4K composers.

\section{Results}
In the following subsections, we present our main results: first, we achieve state-of-the-art quality on LCA-large with much less extensive repository-level pretraining; second, we demonstrate the impact of the context composer choice on the result of repository-level pretraining. 
Additionally, we provide a more detailed comparative study of repository-level pretraining in the Appendix.

\subsection{Benchmarking Against State-of-the-Art}
To estimate the effectiveness of our trained models, we compare them to DeepSeek Coder 1.3B, OpenCoder 1.5B with no repository-level pretraining, and Qwen2.5-Coder 0.5B and 1.5B. These models serve as strong baselines, representing state-of-the-art performance in similar parameter ranges. Results are shown in Table \ref{tab:sota-comparison}. 

\begin{table}[t]
\caption{ 
    Comparison of existing models on LCA-large for the line categories: \textit{inproject} and \textit{infile}. \ FL-4K and PD-16K report Exact Match scores for File-level and Path Distance \texttt{.py} evaluation composers. \ RCB represents the repository-context boost score. 
}
\label{tab:sota-comparison}
\begin{center}
\begin{tabular}{l ccc c ccc}
\toprule
 & \multicolumn{3}{c}{\bf inproject} & & \multicolumn{3}{c}{\bf infile} \\\cmidrule(lr){2-4}\cmidrule(lr){6-8}
 \bf Model               & FL-4K   & PD-16K & RCB & & FL-4K & PD-16K & RCB \\ 
\midrule
Qwen2.5-Coder 0.5B       & 22.6 & 44.2 & +21.6  & & 27.5 & 43.2 & +15.7 \\\cmidrule(lr){1-1} \cmidrule(lr){2-4}\cmidrule(lr){6-8}
DeepSeek Coder 1.3B      & 25.1 & 42.3 & +17.2 & & 30.3 & 43.8 & +13.5 \\
OpenCoder 1.5B           & \underline{26.4} & 0.0 & --26.4   & & 32.6 & 0.0 & --32.6 \\
Qwen2.5-Coder 1.5B       & \textbf{27.2} & \underline{48.5} & \underline{+21.3} & & \textbf{34.3} & \textbf{49.7} & \textbf{+15.4} \\ 

\textbf{Ours} (OpenCoder 1.5B)         &  & & & & \\
~~~File-level pretr.      & 25.9 & 45.2 & +19.3 & & 33.0 & 44.6 & +11.6 \\ 
~~~Path Distance \texttt{.py} pretr.   & 26.2 & \textbf{48.8} & \textbf{+22.6} & & \underline{33.1} & \underline{47.6} & \underline{+14.5} \\
\bottomrule
\end{tabular}
\end{center}
\end{table}

We started with OpenCoder model which is pretty good on file-level code completion among the similar size models and got a significant gain by file-level pretraining on just 1B tokens \footnote{While the model had access to 1B tokens, it was actually used just 72M tokens for training.}. This approach serves as a guideline for scenarios with limited data and low GPU resources, since we do not need repositories, and do not actually need long context for training. We can even achieve Qwen2.5-Coder performance level with 1B tokens of curated repository-level data.

\subsection{Impact of Context Composer Choice}
Findings in the previous subsection leave an open question if there is even better composer for repository-level pretraining.
To answer this question, we evaluate all studied composers and present condensed results for in Table \ref{tab:condensed-results-composers} and extended results in Table \ref{tab:extended-results}. Our experiments demonstrate the performance variations across repository level pretrainings with different context composers.

\begin{table}
\centering
\caption{ 
    Results of evaluating checkpoints after repository-level pretraining. 
    Evaluation dataset is LCA-large for the line categories: \textit{inproject} and \textit{infile}.
    FL-4K and PD-16K report Exact Match scores for File-level and Path Distance \texttt{.py} evaluation composers.
}
\label{tab:condensed-results-composers}
\begin{tabular}{l cc c cc}

\toprule

\multirow{3}{*}{\makecell{\textbf{Pretraining} \\ \textbf{Composer}}}  & \multicolumn{2}{c}{\bf inproject} & & \multicolumn{2}{c}{\bf infile} \\\cmidrule(lr){2-3}\cmidrule(lr){5-6}
& FL-4K & PD-16K & & FL-4K & PD-16K \\

\midrule
Base model (no training) & 26.4 & 0.0 & & 32.6 & 0.0   \\


File-level & 25.9 & 45.2 & & 33.0 & 44.6 \\


Path Distance \texttt{.py} & 26.2 & 48.8 & & 33.1 & 47.6 \\
Other Pretraining Composers & 25.5 -- 26.5 & 46.8 -- 48.7 & & 32.3 -- 33.3 & 45.6 -- 47.8\\





\bottomrule
\end{tabular}
\end{table}

We observe that file-level composer pretraining results in +19.3 repository-context boost, with other pretraining strategies getting repository-context boost within a +20.3 to +22.9 range. Combining with comparable values of the Exact Match, we validate that adapting to the longer context window, \ie new RoPE's base frequency, rather than the specific sequence composition, is the primary factor in repository-level pretraining, with context composers contributing only marginally for suggested approaches.

\section{Conclusion}
In this paper, we address the challenge of project-level code completion by evaluating pretraining for code LLM on various data extracted from repository. 
Our extensive experiments demonstrate that even relatively small training dataset and simple context composer (\eg file-level or path distance) is enough to get a model comparable to the latest state-of-the-art code LLMs. 
This insight reduces the complexity of repository-level pretraining, which effectively minimizes the technical complexities and encourages to broadly research the topic.

Although our findings are promising, they have certain limitations. Our experiments are limited to the OpenCoder model, and it remains unclear whether they generalize to other LLMs. A key direction for future work is to apply our approach on a broader range of Code LLMs. However, recent Code LLMs were released after the repository-level pretraining stage, which may introduce inconsistencies in evaluation.

\subsubsection*{Acknowledgments}
We appreciate the contributions of Alexander Bezzubov, Egor Bogomolov, Timofey Bryksin and Yaroslav Golubev from JetBrains, whose guidance greatly enhanced this research.

\bibliography{iclr2025_conference}
\bibliographystyle{iclr2025_conference}

\newpage
\appendix
\section{Context Composers}
\subsection{Complete List}

All composers follow two standard preprocessing steps, filtering out empty files and normalizing all line separators to Line Feed (LF). With these shared characteristics, the full list of context composers ensures comprehensive coverage for research exploration.

\label{appendix:context-composers}
\begin{enumerate}
    \item \textbf{File-level} — Produces an empty context.
    
    \item \textbf{Path Distance \texttt{.py}} — Constructs the context using only files with the \texttt{.py} extension. The selected files are sorted in descending order based on their path distance from the completion file. If multiple files share the same path distance, a secondary sorting step is applied using the Intersection over Union (IoU) metric, computed over lines shared with the completion file. The IoU metric is calculated on lines with leading and trailing whitespace characters removed, considering only those lines that are at least five characters long after the whitespace removal.
    
    \item \textbf{Lines IoU \texttt{.py}} — Similar to the Path Distance \texttt{.py} method but does not apply the primary sorting step based on path distance. Instead, files are directly ranked using the IoU metric.
    
    \item \textbf{Code Chunks} — Removes all docstrings, comments, and import statements from the context produced by Path Distance \texttt{.py}.
    
    \item \textbf{Half-memory \texttt{.py}} — Starts with the context produced by Path Distance \texttt{.py}. Each line is independently removed with a dropout probability of $0.5$, maintaining the overall saturation of the context window.
    
    \item \textbf{Declarations \texttt{.py}} — Builds upon Path Distance \texttt{.py} by filtering out all non-declarative elements, retaining only function and class declarations.
    
    \item \textbf{Text Chunks \texttt{.py}} — Uses Path Distance \texttt{.py} as the base method. All code is removed from the context, leaving docstrings and comments only.
    
    \item \textbf{Text files} — Constructs the context using files with the extensions \texttt{.json}, \texttt{.yaml}, \texttt{.yml}, \texttt{.sh}, \texttt{.md}, \texttt{.txt}, and \texttt{.rst}. The selected files are grouped in ascending order of relevance: [\texttt{.json}], [\texttt{.yaml}, \texttt{.yml}], [\texttt{.sh}], [\texttt{.md}, \texttt{.txt}, \texttt{.rst}]. Within each group, a secondary sorting step is performed in descending order based on path distance from the completion file.
    
    \item \textbf{Random files} — Constructs the context by randomly ordering all files from the repository snapshot.
    
    \item \textbf{Random \texttt{.py}} — Selects only files with the \texttt{.py} extension and orders them randomly.

    \item \textbf{Mixed context\footnote{Duplication composer is disabled in evaluation mode}} — The context for each data point is constructed by randomly selecting one of the following composers: File-level, Path Distance \texttt{.py}, Half-memory \texttt{.py}, Declarations \texttt{.py}, Text files, Random files, or Duplication.

\end{enumerate}

Furthermore, we propose four additional context composers, which are omitted from the results discussion as they do not reflect realistic scenarios.
\begin{enumerate}    
    \item \textbf{Random tokens} — Constructs the context using a randomly sampled sequence of non-special tokens, each selected independently and with equal probability.
    
    \item \textbf{Duplication} — Constructs the context by concatenating the content of the completion file repeatedly until the maximum context window size is reached.
    
    \item \textbf{Leak} — Starts with the context produced by Path Distance \texttt{.py}. The completion file is randomly split into five segments at newline characters, which then disjointedly replace context lines at random positions, approximately preserving the original token count.

    \item \textbf{Masked Leak} — Starts with the context produced by Path Distance \texttt{.py}. The completion file is divided into segments, each consisting of five lines with one overlapping line at the beginning and one at the end. These segments independently and disjointedly replace context lines at random positions. Additionally, each token in the context has a $0.15$ probability of being replaced with a different non-special token.
\end{enumerate}

For each composer we also consider two modifications:
\begin{itemize}
    \item \textit{reversed} --- we retrieve files that fit into the context window with the composer and reverse their order, so the most relevant one is in the beginning of the context string;
    \item \textit{irrelevant} --- we reverse the order of files obtained from the composer and therefore retrieve most irrelevant files.
\end{itemize}

\subsection{Input Formatting}

All composers employ a uniform strategy for input formatting. The files processed by a composer undergo a predefined formatting pattern \ref{fig:file_representation}, which uses a special token from the OpenCoder's vocabulary. Subsequently, the processed files are concatenated into a single string, referred to as the composed context.

\begin{figure}[H]
    \centering
    \begin{tcolorbox}[colback=white, colframe=black, boxrule=0.5pt, width=0.8\linewidth, sharp corners]
        \centering
        \texttt{<file\_sep>\# \{file\_name\}\textbackslash n\{file\_content\}} \\
    \end{tcolorbox}
    \caption{File Representation}
    \label{fig:file_representation}
\end{figure}

A similar transformation is applied independently to the completion file.

\section{Training Dataset}
\subsection{Filtering}
\label{appendix:data-filtering}
To avoid training on test data, we exclude repositories used in the Long Code Arena's Project-level code completion task. In addition, to ensure data relevance and quality, we apply the following filtering criteria. First, all commits made prior to $2010$ are excluded. Second, completion files with lengths outside the closed interval $[800,25000]$ characters are removed. Third, to eliminate redundancy, a simple deduplication strategy is employed on completion files based on the file name and the name of the repository to which they belong. Finally, up to $1000$ of the most recently updated unique completion files are selected from each repository. The remaining repository snapshot is retained without additional processing.

\section{Training Details}
\label{appendix:training}

\subsection{Hyperparameters}
\label{appendix:training-hyperparams}
The optimization process was conducted using the AdamW optimizer with $\beta_1=0.9$, $\beta_2 = 0.999$, and a weight decay of $0.01$. A batch size of $128$ was employed, with a micro-batch size of $1$ to accommodate hardware constraints. To ensure stable training, gradient clipping was applied with a maximum gradient Euclidean norm of $2$. The learning rate was managed using a cosine decay scheduler with a linear warm-up phase, where the maximum learning rate was set to $5 \times 10^{-5}$. The warm-up phase lasted for $256$ iterations, after which the learning rate followed a cosine decay schedule for $3244$ additional iterations, reaching a minimum value of $5 \times 10^{-8}$.

\subsection{Training Mode of Context Composers}
\label{appendix:training-cococo} 
For training, we obtain an input sequence from each row of the composed dataset by independently tokenizing the context string and the completion file. This process ensures that the completion sequence does not exceed 4,096 tokens and that the total length of the concatenated input remains within 16,384 tokens. To enforce these constraints, we apply truncation from the left for the context and from the right for the completion. Given that most composed contexts exhibit high token saturation, we maintain a context-to-completion token ratio exceeding $3:1$.

\section{Evaluation Details}
\label{appendix:evaluation}

\subsection{Evaluation Mode of Context Composers}
\label{appendix:evaluation-cococo}
For evaluation, we obtain an input sequence for each row of the composed dataset by tokenizing the concatenation of the context string and the completion file. We then apply truncation from the left. 

Compared to the training mode (see Appendix \ref{appendix:training-cococo}), we do not fix the maximum sequence length. Instead, we treat it as a parameter that can be adjusted based on the evaluation requirements. For example, in Appendix \ref{appendix:context-scaling}, we demonstrate the dependency between checkpoint quality and maximum sequence length.

We use the following four evaluation composers in our tables:
\begin{itemize}
    \item \textbf{FL-4K}: File-Level composer with maximum sequence length 4K tokens. We use it to estimate how hard the task is without any repository-context, and as a reference point for calculating gains.
    \item \textbf{PD-4K}: Path Distance composer with maximum sequence length 4K tokens. We use it to estimate model's in-context learning capabilities with initial input sequence length (4K tokens).
    \item \textbf{PD-16K}: Path Distance composer with maximum sequence length 16K tokens. We use it to estimate model's in-context learning capabilities with new input sequence length (16K tokens), and this is the main composer to compare repository-level pretraining with different context composers.
    \item \textbf{Or-16K}: original pretraining composer in evaluation mode with maximum sequence length 16K tokens. We use it to identify the most promising composer overall. This composer applies only to our checkpoints.
\end{itemize}


\section{Comprehensive Compilation of Evaluation Results}

We present results of our experiments in Table \ref{tab:extended-results}. They can be used as baselines for further research.

We additionally include results for the base model with RoPE's base frequencies ($\theta$) being 10,000 and 500,000, results for pretraining with file-level composer for the same values of $\theta$. These results demonstrate that RoPE adjustments impact model quality, and that the model with initial base frequency performs on zero-level for long contexts even after finetuning.

\begin{table}
\centering

\caption{Results of evaluating all checkpoints after repository-level pretraining on all evaluation composers.
\label{tab:extended-results}
Evaluation dataset is LCA-large for the line categories: \textit{inproject} and \textit{infile}.
Highlighted column is the main column for in-context learning capabilities comparison.}

\resizebox{\textwidth}{!}{
    \begin{tabular}{lc cc>{\columncolor{gray!30}}cc c cc>{\columncolor{gray!30}}cc}
    \toprule
    
    \multirow{3}{*}{\makecell{\textbf{Pretraining} \\ \textbf{Composer}}} & & \multicolumn{4}{c}{\bf inproject} & & \multicolumn{4}{c}{\bf infile} \\\cmidrule(lr){3-6}\cmidrule(lr){8-11}
    & & FL-4K & PD-4K & PD-16K & Or-16K & & FL-4K & PD-4K & PD-16K & Or-16K \\
    
    \midrule
    Base model (no training) & & & & & & & & & & \\
    ~~~$\theta = 10{,}000$ & & 26.4 & 36.6 & 0.0 & --- & & 32.6 & 38.2 & 0.0 & --- \\
    ~~~$\theta = 500{,}000$ & & 13.5 & 16.6 & 9.8 & --- & & 15.5 & 12.9 & 4.5 & --- \\
    
    \cmidrule(lr){1-1}
    File-level $4$K & & & & & & & & & & \\
    ~~~$\theta = 10{,}000$ & & 26.2 & 36.4 & 0.0 & 26.2 & & 32.7 & 38.1 & 0.0 & 32.7 \\
    ~~~$\theta = 500{,}000$ & & 25.9 & 36.1 & 45.2 & 25.9 & & 33.0 & 38.1 & 44.6 & 33.0 \\
    
    \cmidrule(lr){1-1}
    Path Distance \texttt{.py} & & 26.2 & 37.0 & 48.8 & 48.8 & & 33.1 & 38.7 & 47.6 & 48.8 \\
    ~~~\textit{reversed} & & 26.1 & 36.9 & 48.3 & 43.2 & & 32.9 & 38.8 & 47.5 & 44.0 \\
    ~~~\textit{irrelevant} & & 25.8 & 36.5 & 47.9 & 26.7 & & 32.5 & 38.1 & 46.7 & 33.4 \\
    
    \cmidrule(lr){1-1}
    Lines IoU \texttt{.py} & & 25.7 & 36.3 & 48.7 & 51.8 & & 33.2 & 38.4 & 47.7 & 50.1 \\
    ~~~\textit{reversed} & & 26.1 & 36.8 & 48.4 & 43.5 & & 33.2 & 38.9 & 47.4 & 44.6 \\
    ~~~\textit{irrelevant} & & 25.8 & 36.4 & 47.5 & 26.7 & & 32.7 & 38.4 & 46.6 & 33.4 \\
    
    \cmidrule(lr){1-1}
    Code Chunks \texttt{.py} & & 25.9 & 36.5 & 47.9 & 47.8 & & 32.8 & 38.2 & 47.5 & 47.9 \\
    ~~~\textit{reversed} & & 26.1 & 36.5 & 47.8 & 41.3 & & 32.8 & 38.3 & 47.4 & 43.0 \\
    ~~~\textit{irrelevant} & & 25.8 & 36.5 & 47.7 & 26.9 & & 32.3 & 37.9 & 46.3 & 33.2 \\
    
    \cmidrule(lr){1-1}
    Half-memory \texttt{.py} & & 25.7 & 36.0 & 47.4 & 38.6 & & 32.9 & 38.4 & 46.6 & 38.7 \\
    ~~~\textit{reversed} & & 25.7 & 36.2 & 47.3 & 35.0 & & 32.9 & 38.2 & 46.5 & 36.9 \\
    ~~~\textit{irrelevant} & & 25.8 & 36.0 & 47.0 & 27.5 & & 32.4 & 37.7 & 46.5 & 33.0 \\
    
    \cmidrule(lr){1-1}
    Declarations \texttt{.py} & & 25.9 & 36.5 & 46.8 & 28.2 & & 32.6 & 38.1 & 46.1 & 34.4 \\
    ~~~\textit{reversed} & & 25.7 & 36.5 & 46.9 & 28.1 & & 32.7 & 38.1 & 45.7 & 34.2 \\
    ~~~\textit{irrelevant} & & 26.2 & 36.3 & 47.2 & 28.2 & & 32.4 & 38.4 & 45.6 & 33.9 \\
    
    \cmidrule(lr){1-1}
    Text Chunks \texttt{.py} & & 26.1 & 36.6 & 47.5 & 26.9 & & 33.0 & 38.5 & 46.9 & 33.2 \\
    ~~~\textit{reversed} & & 26.0 & 36.1 & 47.4 & 26.8 & & 32.9 & 38.5 & 46.2 & 33.5 \\
    ~~~\textit{irrelevant} & & 25.9 & 36.4 & 47.2 & 26.8 & & 32.7 & 38.5 & 46.2 & 33.8 \\
    
    \cmidrule(lr){1-1}
    Text files & & 25.9 & 36.2 & 47.1 & 26.9 & & 33.0 & 38.6 & 46.4 & 33.5 \\
    ~~~\textit{reversed} & & 26.0 & 36.5 & 46.9 & 26.7 & & 33.2 & 38.4 & 46.2 & 33.1 \\
    ~~~\textit{irrelevant} & & 26.0 & 36.5 & 47.1 & 27.0 & & 32.7 & 38.2 & 46.3 & 33.7 \\
    
    \cmidrule(lr){1-1}
    Random files & & 26.2 & 37.0 & 48.1 & 29.8 & & 32.8 & 38.3 & 47.3 & 34.2 \\
    
    \cmidrule(lr){1-1}
    Random \texttt{.py} & & 25.9 & 36.8 & 48.4 & 31.9 & & 32.8 & 38.1 & 47.0 & 35.3 \\

    \cmidrule(lr){1-1}
    Mixed context & & 26.2 & 36.7 & 48.5 & 31.0 & & 32.6 & 38.2 & 47.5 & 36.4 \\
    
    \midrule
    Random tokens & & 26.0 & 36.2 & 44.5 & 26.0 & & 32.6 & 37.9 & 45.1 & 33.1 \\
    
    \cmidrule(lr){1-1}
    Duplication & & 19.6 & 28.8 & 34.7 & 96.7 & & 24.5 & 27.0 & 28.1 & 95.0 \\
    
    \cmidrule(lr){1-1}
    Leak & & 24.9 & 34.8 & 46.1 & 82.9 & & 30.8 & 35.5 & 43.6 & 81.6 \\
    ~~~\textit{reversed} & & 24.3 & 34.7 & 45.6 & 83.8 & & 30.8 & 35.3 & 42.8 & 81.0 \\
    ~~~\textit{irrelevant} & & 24.5 & 34.6 & 45.6 & 82.2 & & 31.3 & 35.6 & 43.2 & 79.7 \\
    
    \cmidrule(lr){1-1}
    Masked Leak & & 25.2 & 35.4 & 46.4 & 65.5 & & 31.6 & 36.9 & 45.0 & 63.5 \\
    
    \bottomrule
    \end{tabular}
}
\end{table}

When using FL-4K composer, the model successfully recovers its quality after RoPE adjustments, suggesting that file-level data alone is sufficient to restore performance. The initial model shows strong in-context learning capabilities for the PD-4K composer, with it outperforming file-level inference. This advantage persists after repository-level pretraining, indicating that training on the collected data effectively retains model's ability to utilize relevant context for shorter context size.

For the PD-16K composer, the initial model, without RoPE adaptation, fails completely, but RoPE scaling alone improves Exact Match scores. Further pretraining yields gains of +19 for file-level training and +22 for the best composer in \textit{inproject} category, with all final scores being slightly higher than file-level pretraining performance. This suggests that adapting to the longer context window, rather than the specific sequence composition, is the primary factor in repository-level code completion, with context composers contributing only marginally for suggested approaches (+3 points for \textit{inproject} category).

Overall, our findings emphasize that RoPE adaptation is the dominant factor in long-context performance gains, while sequence composition plays a secondary role. Future work should explore more effective retrieval-based strategies to maximize repository-level context utilization.

\section{Performance Scaling Beyond Training Context Window}
\label{appendix:context-scaling}

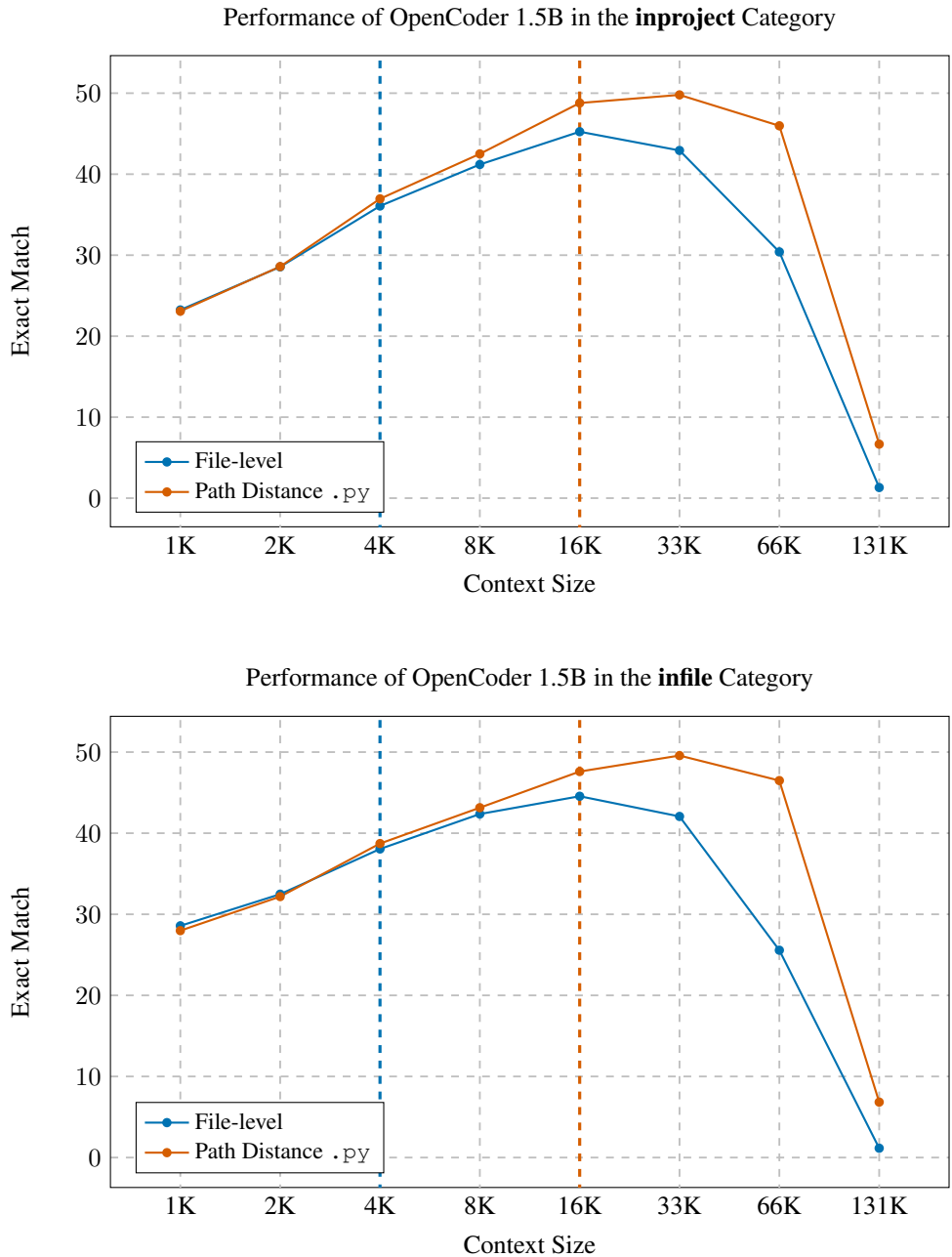
\begin{figure}
\begin{tikzpicture}

\definecolor{custom_blue}{RGB}{0, 114, 178}
\definecolor{custom_orange}{RGB}{213, 94, 0}

\begin{axis}[
    xlabel=Context Size,
    ylabel=Exact Match,
    xmode=log,
    log basis x=2,
    grid style={dashed, line width=0.7pt},
    ymajorgrids=true,
    legend pos=south west,
    width=13cm,
    height=8cm,
    at={(0,9cm)},
    xtick={1024,2048,4096,8192,16384,32768,65536,131072},
    xticklabels={1K,2K,4K,8K,16K,33K,66K,131K},
    tick style={draw=none},
    legend style={font=\small},
    legend cell align={left},
    title=Performance of OpenCoder 1.5B in the \textbf{inproject} Category
]
\draw[dashed, color=custom_blue, line width=1.2pt] (axis cs:4096,-10) -- (axis cs:4096,100);
\draw[dashed, color=custom_orange, line width=1.2pt] (axis cs:16384,-10) -- (axis cs:16384,100);

\draw[dashed, color=black!25, line width=0.7pt] (axis cs:1024,-10) -- (axis cs:1024,100);
\draw[dashed, color=black!25, line width=0.7pt] (axis cs:2048,-10) -- (axis cs:2048,100);
\draw[dashed, color=black!25, line width=0.7pt] (axis cs:8192,-10) -- (axis cs:8192,100);
\draw[dashed, color=black!25, line width=0.7pt] (axis cs:32768,-10) -- (axis cs:32768,100);
\draw[dashed, color=black!25, line width=0.7pt] (axis cs:65536,-10) -- (axis cs:65536,100);
\draw[dashed, color=black!25, line width=0.7pt] (axis cs:131072,-10) -- (axis cs:131072,100);

\addplot[
    color=custom_blue,
    mark=*,
    mark options={
        scale=0.7,
    },
    thick,
] coordinates {
    (1024,23.236994)
    (2048,28.554913)
    (4096,36.069364)
    (8192,41.194605)
    (16384,45.240848)
    (32768,42.928709)
    (65536,30.404624)
    (131072,1.310212)
};

\addplot[
    color=custom_orange,
    mark=*,
    mark options={
        scale=0.7,
    },
    thick,
] coordinates {
    (1024,23.082852)
    (2048,28.593449)
    (4096,36.955684)
    (8192,42.504817)
    (16384,48.786127)
    (32768,49.788054)
    (65536,45.973025)
    (131072,6.666667)
};
\legend{
    {File-level},
    {Path Distance \texttt{.py}}
}
\end{axis}

\begin{axis}[
    xlabel=Context Size,
    ylabel=Exact Match,
    xmode=log,
    log basis x=2,
    grid style={dashed, line width=0.7pt},
    ymajorgrids=true,
    legend pos=south west,
    width=13cm,
    height=8cm,
    xtick={1024,2048,4096,8192,16384,32768,65536,131072},
    xticklabels={1K,2K,4K,8K,16K,33K,66K,131K},
    tick style={draw=none},
    legend style={font=\small},
    legend cell align={left},
    title=Performance of OpenCoder 1.5B in the \textbf{infile} Category
]

\draw[dashed, color=custom_blue, line width=1.2pt] (axis cs:4096,-10) -- (axis cs:4096,100);
\draw[dashed, color=custom_orange, line width=1.2pt] (axis cs:16384,-10) -- (axis cs:16384,100);

\draw[dashed, color=black!25, line width=0.7pt] (axis cs:1024,-10) -- (axis cs:1024,100);
\draw[dashed, color=black!25, line width=0.7pt] (axis cs:2048,-10) -- (axis cs:2048,100);
\draw[dashed, color=black!25, line width=0.7pt] (axis cs:8192,-10) -- (axis cs:8192,100);
\draw[dashed, color=black!25, line width=0.7pt] (axis cs:32768,-10) -- (axis cs:32768,100);
\draw[dashed, color=black!25, line width=0.7pt] (axis cs:65536,-10) -- (axis cs:65536,100);
\draw[dashed, color=black!25, line width=0.7pt] (axis cs:131072,-10) -- (axis cs:131072,100);

\addplot[
    color=custom_blue,
    mark=*,
    mark options={
        scale=0.7,
    },
    thick,
] coordinates {
    (1024,28.576737)
    (2048,32.478632)
    (4096,38.052768)
    (8192,42.363434)
    (16384,44.555927)
    (32768,42.066146)
    (65536,25.566704)
    (131072,1.151988)
};

\addplot[
    color=custom_orange,
    mark=*,
    mark options={
        scale=0.7,
    },
    thick,
] coordinates {
    (1024,27.982163)
    (2048,32.181345)
    (4096,38.721665)
    (8192,43.143813)
    (16384,47.603122)
    (32768,49.572650)
    (65536,46.488294)
    (131072,6.837607)
};
\legend{
    {File-level},
    {Path Distance \texttt{.py}}
}
\end{axis}
\end{tikzpicture}

\caption{Performance comparison of File-level and Path Distance \texttt{.py} approaches across different context sizes for OpenCoder 1.5B model. The plots show the Exact Match accuracy for both inproject (top) and infile (bottom) categories. The dashed vertical lines represent the context length used during repository-level pretraining.}
\label{fig:beyound_training_context_window}
\end{figure}

The repository-level pretraining with File-level composer and Path Distance composer for maximum sequence lengths of 4K and 16K. However, pretrained checkpoints extrapolate beyond these lengths up to 16K and 32K as snown on Figure \ref{fig:beyound_training_context_window}.

Similar behavior was observed in \citet{codellama} and it needs additional research.

\section{Masked Loss and Full Loss Results}
\label{appendix:masked-vs-full}

Some context composers create out-of-distribution sequences (\eg Declarations \texttt{.py}). We avoid distribution shift by masking the loss, \ie use only gradients from completion file tokens for training. In case of any composer that includes unprocessed code files in the context, we lose tokens for training. However, the results for each composer in Table \ref{tab:masked-vs-full-loss} are comparable for masked loss and full loss pretraining, with the only exception being the Duplication composer.

\begin{table}[h]
\centering
\caption{Comparison of checkpoints pretrained with masked loss and full loss.}
\label{tab:masked-vs-full-loss}
\resizebox{\textwidth}{!}{
    \begin{tabular}{lc cccc c cccc}
    \toprule
    
    \multirow{3}{*}{\makecell{\textbf{Pretraining} \\ \textbf{Composer}}} & & \multicolumn{4}{c}{\bf inproject} & & \multicolumn{4}{c}{\bf infile} \\\cmidrule(lr){3-6}\cmidrule(lr){8-11}
    & & FL-4K & PD-4K & PD-16K & Or-16K & & FL-4K & PD-4K & PD-16K & Or-16K \\
    
    \midrule
    Path Distance \texttt{.py} & & & & & & & & & & \\
    ~~~Masked loss & & 26.2 & 37.0 & 48.8 & 48.8 & & 33.1 & 38.7 & 47.6 & 47.6 \\
    ~~~Full loss & & 26.3 & 36.5 & 48.4 & 48.4 & & 33.1 & 38.6 & 47.8 & 47.8 \\
    
    \midrule
    Path Distance \texttt{.py}, \textit{reversed} & & & & & & & & & & \\
    ~~~Masked loss & & 26.1 & 36.9 & 48.3 & 43.2 & & 32.9 & 38.8 & 47.5 & 44.0 \\
    ~~~Full loss & & 26.2 & 36.8 & 48.4 & 43.1 & & 33.1 & 38.6 & 47.2 & 44.1 \\
    
    \midrule
    Path Distance \texttt{.py}, \textit{irrelevant} & & & & & & & & & & \\
    ~~~Masked loss & & 25.8 & 36.5 & 47.9 & 26.7 & & 32.5 & 38.1 & 46.7 & 33.4 \\
    ~~~Full loss & & 25.5 & 36.5 & 48.0 & 26.5 & & 33.0 & 38.4 & 46.8 & 33.6 \\
    
    \midrule
    Lines IoU \texttt{.py} & & & & & & & & & & \\
    ~~~Masked loss & & 25.7 & 36.3 & 48.7 & 51.8 & & 33.2 & 38.4 & 47.7 & 50.1 \\
    ~~~Full loss & & 25.9 & 36.6 & 48.4 & 51.2 & & 32.9 & 38.8 & 47.5 & 49.6 \\
    
    \midrule
    Lines IoU \texttt{.py}, \textit{reversed} & & & & & & & & & & \\
    ~~~Masked loss & & 26.1 & 36.8 & 48.4 & 43.5 & & 33.2 & 38.9 & 47.4 & 44.6 \\
    ~~~Full loss & & 26.0 & 36.8 & 48.3 & 43.6 & & 33.1 & 38.9 & 47.4 & 44.8 \\
    
    \midrule
    Lines IoU \texttt{.py}, \textit{irrelevant} & & & & & & & & & & \\
    ~~~Masked loss & & 25.8 & 36.4 & 47.5 & 26.7 & & 32.7 & 38.4 & 46.6 & 33.4 \\
    ~~~Full loss & & 26.2 & 36.8 & 48.2 & 26.6 & & 33.3 & 38.7 & 47.2 & 33.3 \\
    
    \midrule
    Code Chunks \texttt{.py} & & & & & & & & & & \\
    ~~~Masked loss & & 25.9 & 36.5 & 47.9 & 47.8 & & 32.8 & 38.2 & 47.5 & 47.9 \\
    ~~~Full loss & & 26.5 & 36.8 & 48.7 & 47.7 & & 33.2 & 38.6 & 47.8 & 47.8 \\
    
    \midrule
    Code Chunks \texttt{.py}, \textit{reversed} & & & & & & & & & & \\
    ~~~Masked loss & & 26.1 & 36.5 & 47.8 & 41.3 & & 32.8 & 38.3 & 47.4 & 43.0 \\
    ~~~Full loss & & 26.4 & 37.0 & 48.7 & 41.5 & & 33.3 & 38.8 & 47.8 & 43.4 \\
    
    \midrule
    Code Chunks \texttt{.py}, \textit{irrelevant} & & & & & & & & & & \\
    ~~~Masked loss & & 25.8 & 36.5 & 47.7 & 26.9 & & 32.3 & 37.9 & 46.3 & 33.2 \\
    ~~~Full loss & & 25.7 & 36.5 & 48.1 & 26.9 & & 33.3 & 38.7 & 47.1 & 33.3 \\
    
    \midrule
    Random \texttt{.py} & & & & & & & & & & \\
    ~~~Masked loss & & 25.9 & 36.8 & 48.4 & 31.9 & & 32.8 & 38.1 & 47.0 & 35.3 \\
    ~~~Full loss & & 26.2 & 36.8 & 48.3 & 32.2 & & 33.1 & 38.5 & 47.6 & 36.0 \\
    
    \midrule
    Duplication & & & & & & & & & & \\
    ~~~Masked loss & & 19.6 & 28.8 & 34.7 & 96.7 & & 24.5 & 27.0 & 28.1 & 95.0 \\
    ~~~Full loss & & 25.5 & 35.9 & 46.4 & 97.3 & & 32.8 & 38.4 & 44.4 & 96.4 \\
    
    \bottomrule
    \end{tabular}
}
\end{table}

\end{document}